\documentclass[useAMS,usenatbib]{mn2e}

\usepackage{psfig}
\usepackage[dvips]{graphicx}
\usepackage{lscape}

\def\lesssim{\mathrel{\hbox{\rlap{\hbox{\lower4pt\hbox{$\sim$}}}\hbox{$<$}}}}
\def\gtrsim{\mathrel{\hbox{\rlap{\hbox{\lower4pt\hbox{$\sim$}}}\hbox{$>$}}}}

\title{The star formation activity in cosmic voids}
\author[Ricciardelli et al.]{E. Ricciardelli$^{1}$\thanks{E-mail:
    elena.ricciardelli@uv.es}, A. Cava$^{2}$, J. Varela$^{3}$, V. Quilis$^{1,4}$\\
$^{1}$Departament d'Astronomia i Astrofisica, Universitat de Valencia, c/ Dr. Moliner 50, E-46100 - Burjassot, Val{\`e}ncia, Spain\\
$^{2}$Observatoire de Gen{\`e}ve, Universit{\'e} de Gen{\`e}ve, 51 Ch. des Maillettes, 1290 Versoix, Switzerland\\
$^{3}$Centro de Estudios de F\'{\i}sica del Cosmos de Arag\'on
(CEFCA), Plaza San Juan 1, 44001 Teruel, Spain\\
$^{4}$Observatori Astron{\`o}mic, Universitat de Val{\`e}ncia, E-46980 Paterna, Val{\`e}ncia, Spain
}

\begin{document}
\date{Accepted ...  Received ...; in original form ...
}
\maketitle
\label{firstpage}
\begin{abstract}

Using a sample of cosmic voids identified in the Sloan Digital Sky
Survey Data Release 7, we study the star formation activity of void
galaxies. The properties of galaxies living in voids are
compared with those of galaxies living in the 
void shells and with a control sample, representing the general galaxy
population. 
Void galaxies appear to form stars more efficiently than shell
galaxies and the control sample. This result can not be interpreted
as a consequence of the bias towards low masses in underdense regions, as void
galaxy subsamples with the same mass distribution as the control
sample also show statistically different specific star formation
rates. This highlights the  fact that galaxy evolution in voids is
slower with respect to the evolution of  the general
population.
Nevertheless, when only the star forming galaxies are considered, we
find that the star formation rate is insensitive to the environment,
as the main sequence is remarkably constant in the three samples under
consideration. 
This fact implies that  environmental effects manifest themselves as
fast quenching mechanisms, while leaving the non-quenched galaxies
almost unaffected,  as  their star formation activity is largely
regulated by the mass of their halo.
We also analyse galaxy properties as a function of void-centric
distance and find that the enhancement in the star formation activity
with respect to the control sample is observable up to a radial
distance $1.5 \cdot R_{void}$. This result  can be used as a suitable
definition of void shells.  
Finally, we find that larger voids show an enhanced star formation activity in
the shells with respect to their smaller counterparts, that could be related to the different dynamical evolution experienced by voids of different sizes.

\end{abstract}
 

\section{Introduction}\label{intro}

Large redshift surveys \citep{York00, Colless01}  and
cosmological simulations \citep{Bond96, AragonCalvo10, Cautun14} have
revealed that galaxies are distributed inside a cosmic web 
of walls, filaments and compact clusters. Such a web encloses large
underdense regions, referred to as cosmic voids. 

Voids were first recognized in the earliest redshift surveys
\citep{Gregory78, Kirshner81}  as huge empty holes in the galaxy distribution. 
Nowadays, there is a general consensus in that voids occupy most of the
volume of the Universe \citep{Sheth04, Vandew11, Pan12} and that they
are far from being simple structures. As shown by
numerical simulations \citep{Vandew93, Sheth04, AragonCalvo13, Ricciardelli13},
voids host a rich infrastructure, made of subvoids, sheet-like structures and tenuous
filaments. These filamentary features have also been observed in the
real voids \citep{Beygu13, Alpaslan14}
and are expected to be the favourite sites for galaxy formation in
voids \citep{Rieder13}.

Voids represent a unique and pristine environment for galaxy formation
studies, since void galaxies are not affected by the transformation
processes (such as ram-pressure stripping, starvation and harassment)
that act in groups and clusters.  Thus, they allow one to study galaxy
evolution as  a result of nature only, in the absence of nurture. 

In the general environment, it is widely known that the star formation
rate (SFR) depends on local density, with galaxies living in dense regions  
having their star formation rates  strongly reduced with respect  to field galaxies
\citep{Balogh97, Poggianti99, Elbaz07}. Evidence has also emerged showing that this SFR-density relation is
largely due to an increasing fraction of passive galaxies in dense
environment, whereas the star-forming population does not show 
any significant trend with the environment \citep{Peng10, Wijesinghe12}. 
This would suggest that any mechanism responsible for the suppression
of the SFR would act on a very short time-scale. 
In the rich clusters, however, there are indications that the SFR of
star forming galaxies also
depends on density \citep{Vulcani10}.

One way to shed light into the effect of environment on galaxy
evolution is to focus on the rarefied void regions, where the
nurture processes are not at work and any environmental trend should
be driven by 'in-situ' processes.
As revealed by observational studies of statistical samples of voids,
galaxies in voids are bluer, have higher specific star formation rates and
are of later types than galaxies living in regions at average
density \citep{Rojas04, Rojas05, Patiri06, vonbenda08, Hoyle12, Kreckel12}.  
A steady increase in the star formation activity down to the densities
typical of voids has been also observed in the surroundings of the Coma
supercluster \citep{Cybulski14}.
However, it is still under debate whether this youthful state of void
galaxies has an intrinsic nature or it is just a consequence of the 
mass bias, since the low-mass galaxies dominate the low-density
environments.
 \citet{Patiri06} find that the colours of void galaxies
are not significantly different than those of wall galaxies of the same
morphological type. Likewise, \citet{Kreckel12}, studying
a sample of 60 void galaxies from the Void Galaxy Survey (VGS, van de Weygaert et al. 2011)
conclude that the void galaxy properties do not differ from those of field
galaxies of the same luminosity and morphology. 
On the other hand,  other works \citep{Rojas04, Hoyle12}
found that the blueness of void galaxies is still
recovered when the galaxy population is divided in morphological
types. Thus, the peculiarities of void galaxies can not be interpreted as a simple consequence of the
morphology-density relation \citep{Dressler80}. If this effect is real, it would
indicate that galaxy evolution in voids is also environmentally driven,
although the environmental effects responsible for galaxy
transformations should be necessarily different than those acting in
high-density regions. 

In this work, we aim at studying the star formation properties of void
galaxies, with the largest sample ever used for this purpose, and to
shed light onto the peculiarities of void galaxies with respect to the
galaxies living in the general environment. We do this by using a
published catalogue of voids \citep{Varela12}, drawn from SDSS and
including more than 6000 void galaxies.
The structure of the paper is as follows. In Section \ref{data} we
introduce our catalogue of void and the galaxy samples used in the
analysis, in Section \ref{results} we present our results and conclude
in Section \ref{conclusions}.

Throughout the paper we adopt the following cosmology: $\Omega_m=0.3$,
$\Omega_{\Lambda}=0.7$ and all the relevant quantities are rescaled to
h=$H_0$/100km/s/Mpc.

\section{The data}\label{data}

\subsection{The SDSS void catalogue}\label{sdss}

The catalogue of cosmic voids used for the present analysis has been
described in \citet{Varela12}. Here, we only give a brief
description of  the void catalogue and refer to the original work for
further details.

The galaxy sample used for void
identification has been extracted from the New York
University Value-Added Galaxy Catalog\footnote{http://sdss.physics.nyu.edu/vagc/} (NYU-VACG; \citealt{Blanton05}), based on the
photometric and spectroscopic catalog of
SDSS/DR7\footnote{http://cas.sdss.org/astrodr7/en}, complete down to
$r\sim 17.77$.  To guarantee the
homogeneity of the sample and avoid the detection of spurious voids, a
catalogue complete down to magnitude
$M_r-5logh=-20.17$ in the redshift range: $0.01 \leq z \leq 0.12$ has
been used. 
Using this galaxy sample, voids are
defined as spherical regions devoid of galaxies.
 As in \citet{Varela12}, we consider only voids whose radius is
larger than $10 \, h^{-1}\,  Mpc$.  The catalogue thus includes 699
voids, which, by definition, can host only galaxies fainter than
$M_r-5log(h)=-20.17$. 

We also define shell galaxies as those galaxies lying at a distance
$\leq 30 \, h^{-1}\,  Mpc$ from the center of a void.  For consistency with the void
galaxy catalogue, in the shell catalogue, we only include galaxies fainter 
 than  $M_r-5log(h)=-20.17$. 
Given the relatively large width of the shell,  it might occur that the same galaxy 
 belongs to the overlapping shells of different voids.  In these cases, and unless otherwise stated, we
associate the given galaxy to the closest void, thus avoiding
multiple occurrences of the same galaxy in the catalog. 

\begin{figure}
\includegraphics[width=\columnwidth]{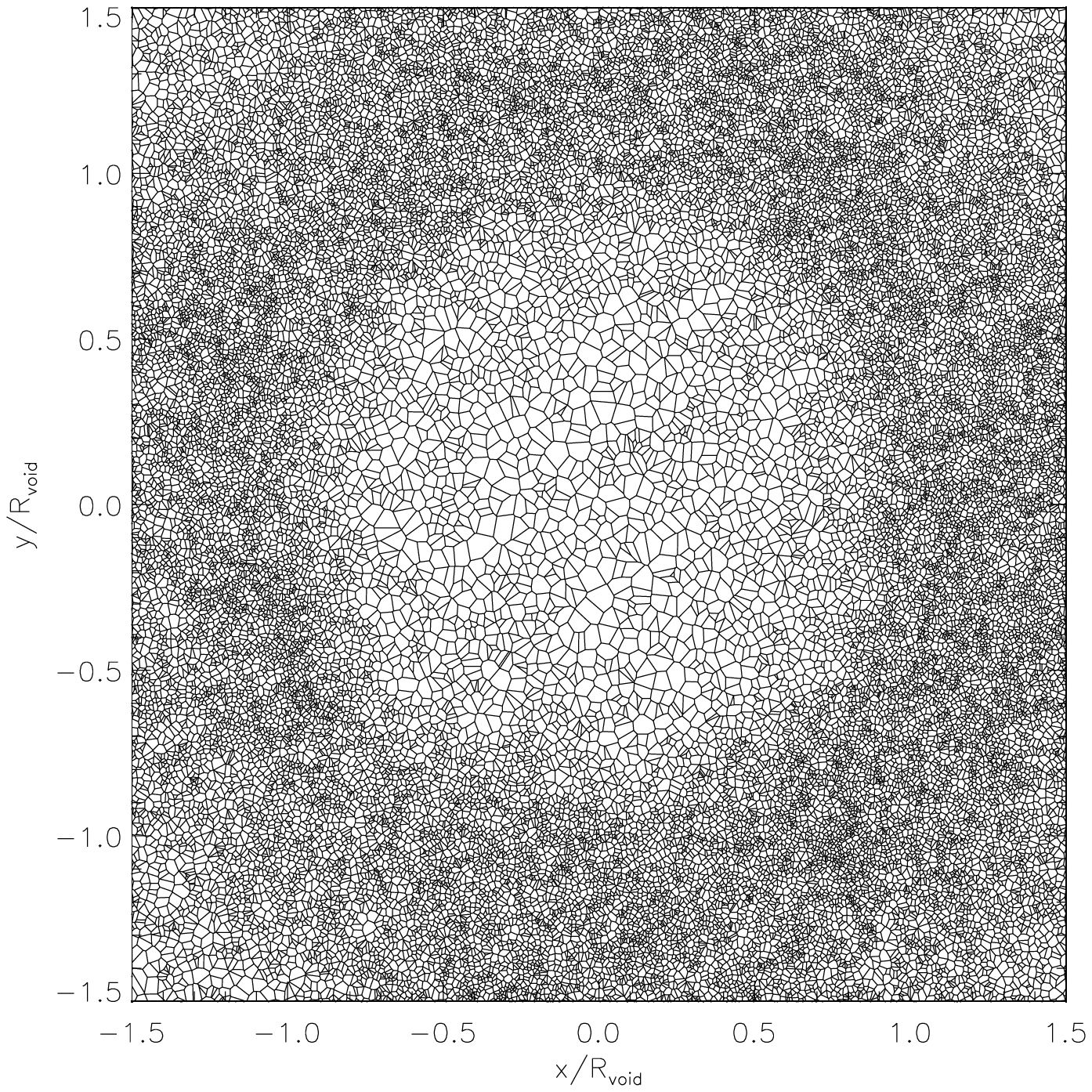}\\
\includegraphics[width=0.95\columnwidth]{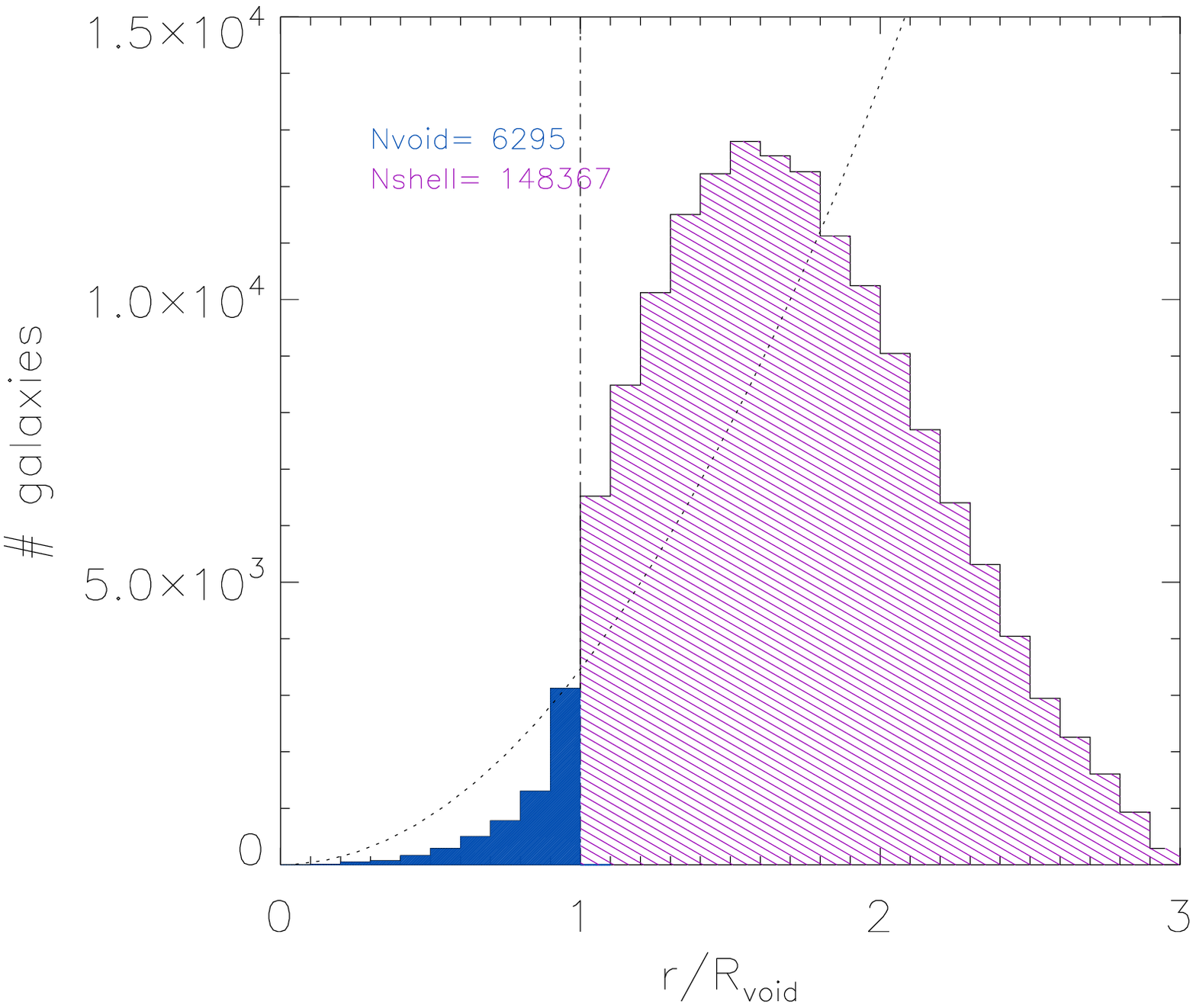}
\caption{{\it Upper panel:}  two-dimensional
representation of the stacked void in comoving coordinates, normalized
to the size of the void.  A slice of thickness 0.5$\cdot R_{void}$ within the void
center has been considered for projection. The
cells represent the Voronoi tessellation of the two-dimensional
distribution of void and shell galaxies. {\it Lower panel:} Number of galaxies as a function of their void-centric
  distance, normalized to the size of the void hosting
  the galaxies. The sharp increase of the number of galaxies at the
  void wall ($r/R_{void} \sim 1$, vertical line) is clearly seen. We also show, for comparison, the number
of galaxies expected for a sphere of constant density, where the
increasing number of galaxies with radius depends only on the increasing volume
of the shells. Here
and in the following plots, blue refers to void galaxies and magenta
indicates the shell galaxies. }
\label{profile} 
\end{figure}

In the upper panel of Fig. \ref{profile} we show the stacked two-dimensional distribution of void and
shell galaxies, considering all the 699 voids of the sample. The
density of galaxies is represented by the Voronoi tessellation of the
galaxy distribution and it clearly shows the lack of galaxies
within $R_{void}$.  The Voronoi tessellation has been computed by
  means of the QHULL procedure in IDL, that constructs convex hulls
  for a 2d distribution of points. 
The number of galaxies as a function of the void-centric distance 
rescaled to the void radius, $r/R_{void}$, is shown in the lower panel
of Fig.  \ref{profile}.  The sharp increase in the number of galaxies towards
the void edge reflects the  shape of the void density profile \citep{Ricciardelli14}. 
For the largest voids in the sample, the physical
distance of  $30 \, h^{-1}\,  Mpc$, used to define the shell galaxies, corresponds
to a rescaled distance $r \sim 1.5 \cdot R_{void}$. Hence, only small voids
contain shells larger than this value, resulting 
in the decaying distribution at  $r > 1.5 \cdot R_{void}$  seen in Fig. \ref{profile}.
In the following, shell galaxies out to $30 \, h^{-1}\,  Mpc$ are
considered only when studying trends with the void-centric
distance (see \ref{dist} and \ref{dist_rvoid}). The rest of the
analysis is restricted to galaxies having $r \leq 1.5 \cdot R_{void}$.

In addition to the void and shell catalogues, we also build a control
sample, including all  galaxies  fainter than
$M_r-5log(h)=-20.17$ and within the same redshift range of the void
sample: $0.01 \le z \le 0.12$. A summary
of  the samples used in this work  is given in Table 1.  

We notice that, due to our definition of voids and void galaxies, we
can study the environmental effect of voids only on  faint galaxies.
However, this is not a limitation to our analysis, as dwarf galaxies
and low mass galaxies in general, are those showing the greatest
peculiarities \citep{Pustilnik11, Pustilnik13, Cybulski14} with
respect to galaxies living in higher density environment. 

\begin{table}
\begin{minipage}{85mm}
\caption{Galaxy samples used in this work. }
\begin{tabular}{cccc}
\hline \hline
{\bf Sample} & {\bf  $N_{total}$}\footnote{$N_{total}$ refers to the
    number of galaxies in the sample  fainter than  $M_r-5log(h)=-20.17$ and with redshift
    in the range $0.01 \leq z \leq 0.12$.}
& {\bf $N_{clean}$}\footnote{\small $N_{clean}$ refers to the number of
  galaxies in the sample, excluding AGN, LINER and {\it unclassifiable} objects} \\
\hline
Void sample & 7210 & 6295 \\
Shell sample ($r \leq  30 \, h^{-1}\,  Mpc$) &171873 & 148367 & \\
Shell sample ($r/R_{void} \leq 1.5$) & 56477 & 48856 \\
Control sample & 225822 & 195222 \\ 
\hline \hline
\label{tab1}
\end{tabular}
\end{minipage}
\end{table} 

\normalsize

\subsection{Star formation rates and stellar masses}

The SFR and stellar mass estimates used in this paper
are those from the MPA
catalogue\footnote{http://www.mpa-garching.mpg.de/SDSS/DR7/}
\citep{Brinchmann04}.  The SFR measurements combine a 
spectroscopic determination of the SFR within the fiber with a
photometric one outside it.  The SFRs within the fiber are derived
from the emission lines, and are
primarily based on the intensity of the $H_{\alpha}$ line.  The
observed spectra are fitted with a grid of template models, which 
combine the stellar population models from \citet{BC03}
with emission line modelling from \citet{CL01}. 
Dust attenuation is also taken into account by considering the
\citet{CF00} extinction law.  The procedure thus produces
dust-corrected SFR.
Since the SDSS spectra come from a 3'' fiber, aperture corrections are
needed to account for the missing flux outside the fiber. 
To estimate it, stellar population models are
used to fit the observed photometry and infer the SFR. The total
SFR is thus given by summing the SFR in the fiber and the one from the photometry. 
The stellar masses presented in the MPA catalogue, and used in this
work, are those measured in \citet{Kauff03}.
SFRs and stellar masses are computed assuming a
\citet{Kroupa01} initial mass function.

In addition, the \citet{Brinchmann04} catalogue  provides a
spectral classification, based on emission lines (BPT diagrams,
\citealt{Baldwin81}),  useful to separate
star-forming galaxies from AGN.
To avoid  having our SFRs   contaminated by the presence of
AGN, we exclude from the samples galaxies classified as AGN or LINER
according to the Brinchmann classification.  We also exclude galaxies
classified as {\it unclassifiable}. In Table 1 we report the
number of galaxies in each sample. In all the
samples, the number of galaxies excluded because AGN, LINER or {\it
  unclassifiable}, contribute by $\sim 13\% $ to the total number of galaxies.

\subsection{Correction for incompleteness}\label{incompleteness}

\begin{figure*}
\includegraphics[width=1.8\columnwidth]{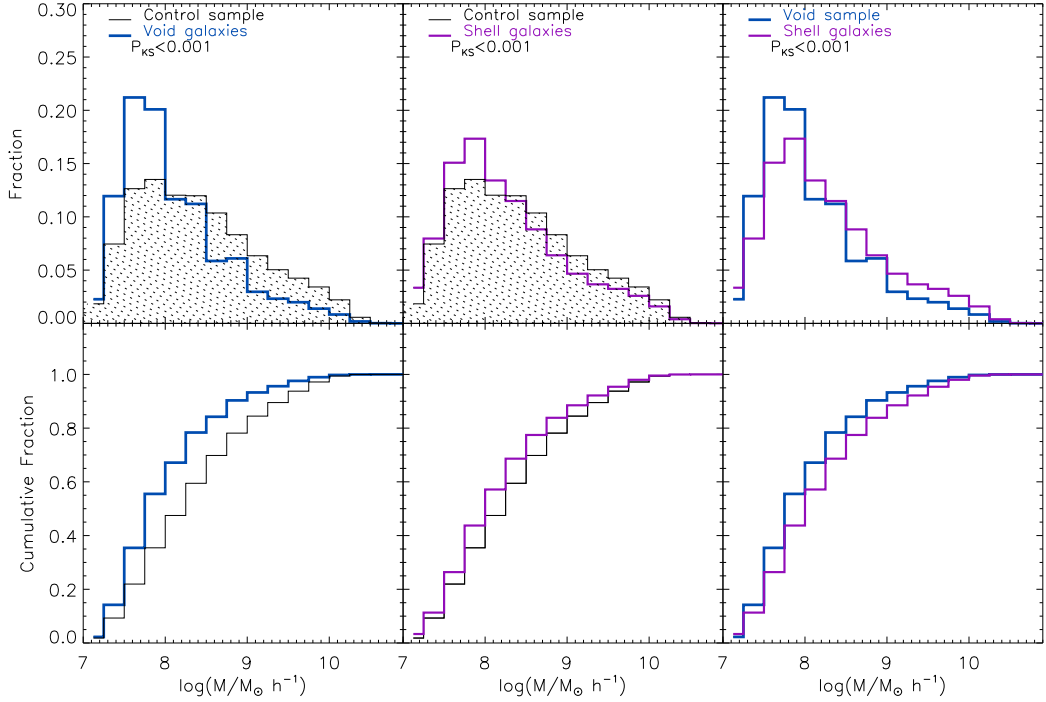}\\
\caption{Upper (lower) panels: differential (cumulative) stellar mass distributions for void galaxies (blue),
  shell galaxies (magenta) and control sample
  (black). In each panel, the probability that the
  two distributions are identical according to the Kolmogorov Smirnov
  test is indicated. The null hypothesis that the two distributions
  are identical can be rejected in all the cases.}
\label{mass_hist} 
\end{figure*} 

The spectroscopic completeness limit of the SDSS DR7 survey 
($m_r<$17.77) implies  a redshift-dependent absolute magnitude limit. 
Hence, our galaxy samples, being fainter than $-20.17+5\log(h)$, are not complete and
we need to correct for this Malmquist bias.
In order to account for this drawback, we correct the sample 
by weighting each galaxy by  $1/V_{max}$ values, where $V_{max}$ is
the volume out to the comoving distance at which the galaxy would
still be observable.

Broadly speaking, we perform the following steps. 
The apparent magnitude limit $m_r=17.77$ of the survey can be
translated into a redshift-dependent absolute magnitude (see Appendix
of \citealt{vdb08}) in the
absolute frame at z=0.1:
\begin{equation}
^{0.1}M_{r}-5\log(h)=m_{r}-DM(z)-k_{0.1}(z)+1.62(z-0.1)-0.1
\end{equation}
where $DM(z)$ is the distance modulus, $k_{0.1}(z)$ is the
K-correction to z=0.1, whose redshift dependence can
be approximated  by \citep{Blanton07}:
\begin{equation}
k_{0.1}(z)=2.5\log\Big(\frac{z+0.9}{1.1}\Big)
\end{equation}
and the term $-0.1$ at the end of the expression takes into account
the scatter in the K-correction.
Hence, for each galaxy in the sample, we consider its rest-frame 
$^{0.1}M_{r}$ from the NYU-VAGC catalogue and compute the maximum
redshift, $z_{max}$ at which the galaxy would be brighter than $m_r=17.77$. The comoving
volume enclosed between z=0.01 and $z_{max}$ gives $V_{max}$.
The weights are finally given by:
\begin{equation}
w_i=\frac{1}{V_{max}C}
\end{equation}
where C is the spatially-dependent spectroscopic completeness factor,
available from the NYU-VAGC.  Thus, $w_i$ gives the number of objects
per unit volume in a complete sample.

\section{Results}\label{results}

\subsection{Distributions in galaxy properties}

\begin{figure*}
\includegraphics[width=1.8\columnwidth]{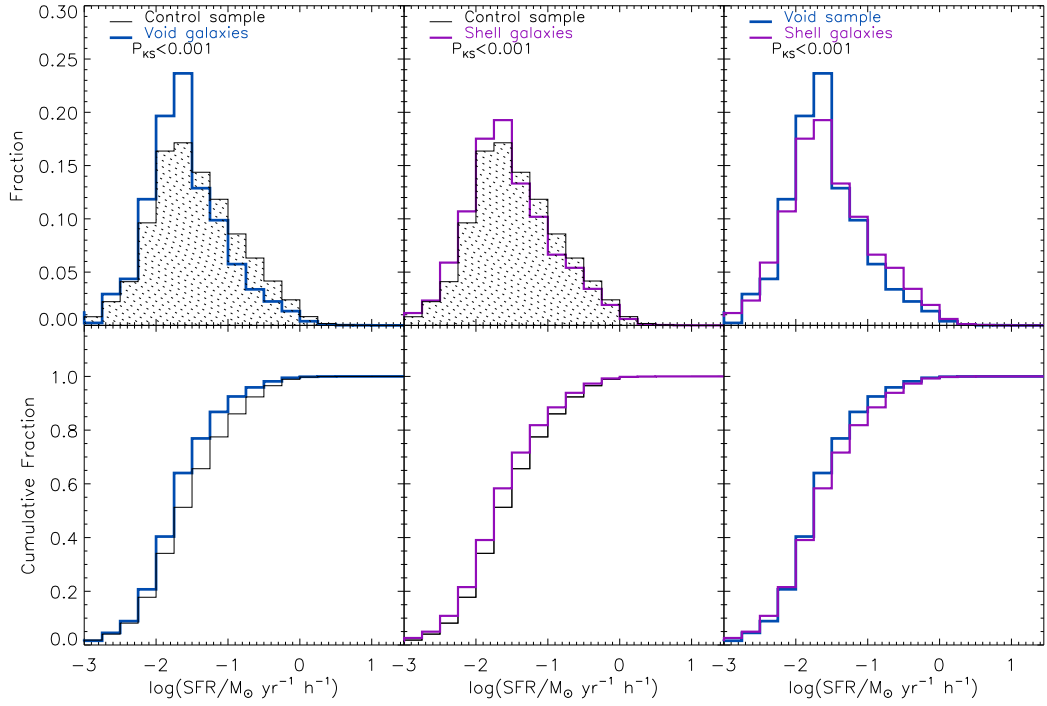}\\
\caption{Differential and cumulative SFR distributions for void galaxies (blue
  histograms), shell galaxies (magenta histograms) and control sample
  (grey histograms). Colors are coded as in Fig. \ref{mass_hist}.}
\label{sfr_hist} 
\end{figure*} 

\begin{figure*}
\includegraphics[width=1.8\columnwidth]{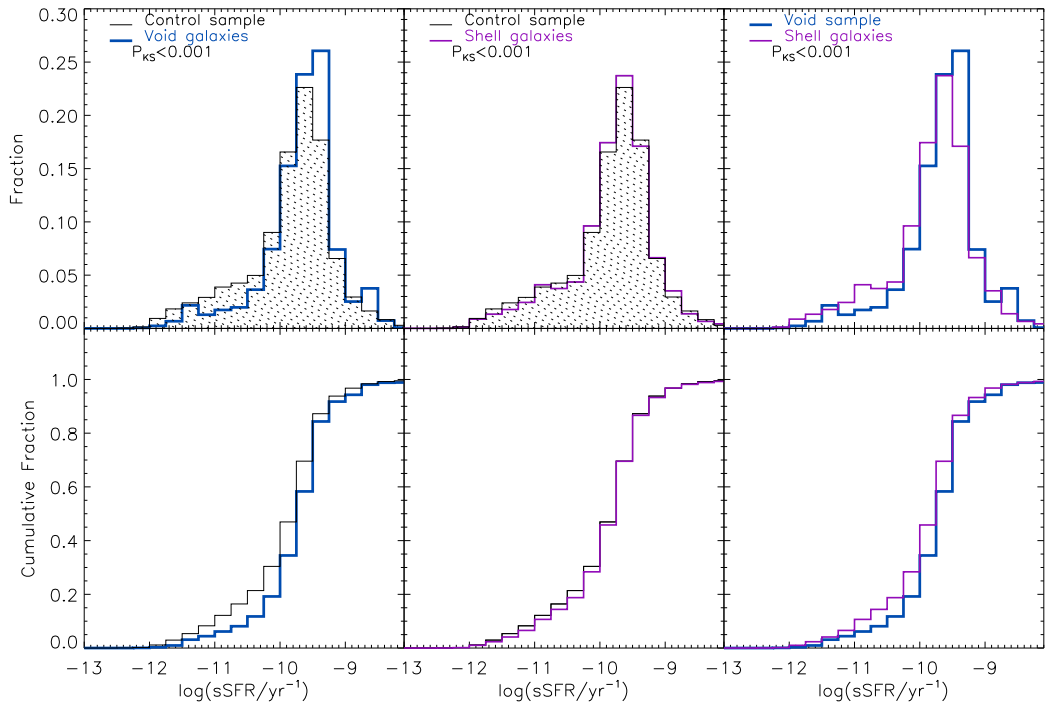}\\
\caption{Differential and cumulative sSFR distributions  for void
  galaxies, shell galaxies and control sample. Colors are coded as in Fig. \ref{mass_hist}.}
\label{ssfr_hist} 
\end{figure*} 

\begin{figure*}
\includegraphics[width=1.8\columnwidth]{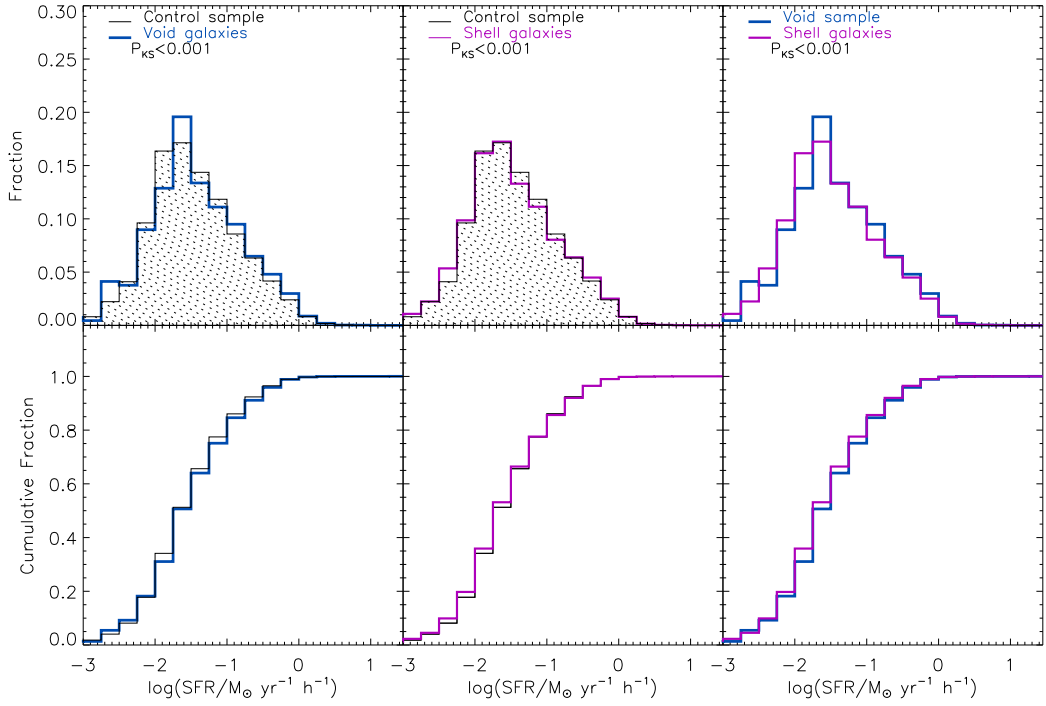}\\
\caption{Differential (upper panels) and cumulative (lower panels)
  SFR distributions of the three samples, using mass-matched samples
  for the void and shell galaxies, in order to avoid the bias induced
  by different mass distributions. See text for further details. Colors are coded as in Fig. \ref{mass_hist}.
}
\label{sfr_hist_m} 
\end{figure*} 

\begin{figure*}
\includegraphics[width=1.8\columnwidth]{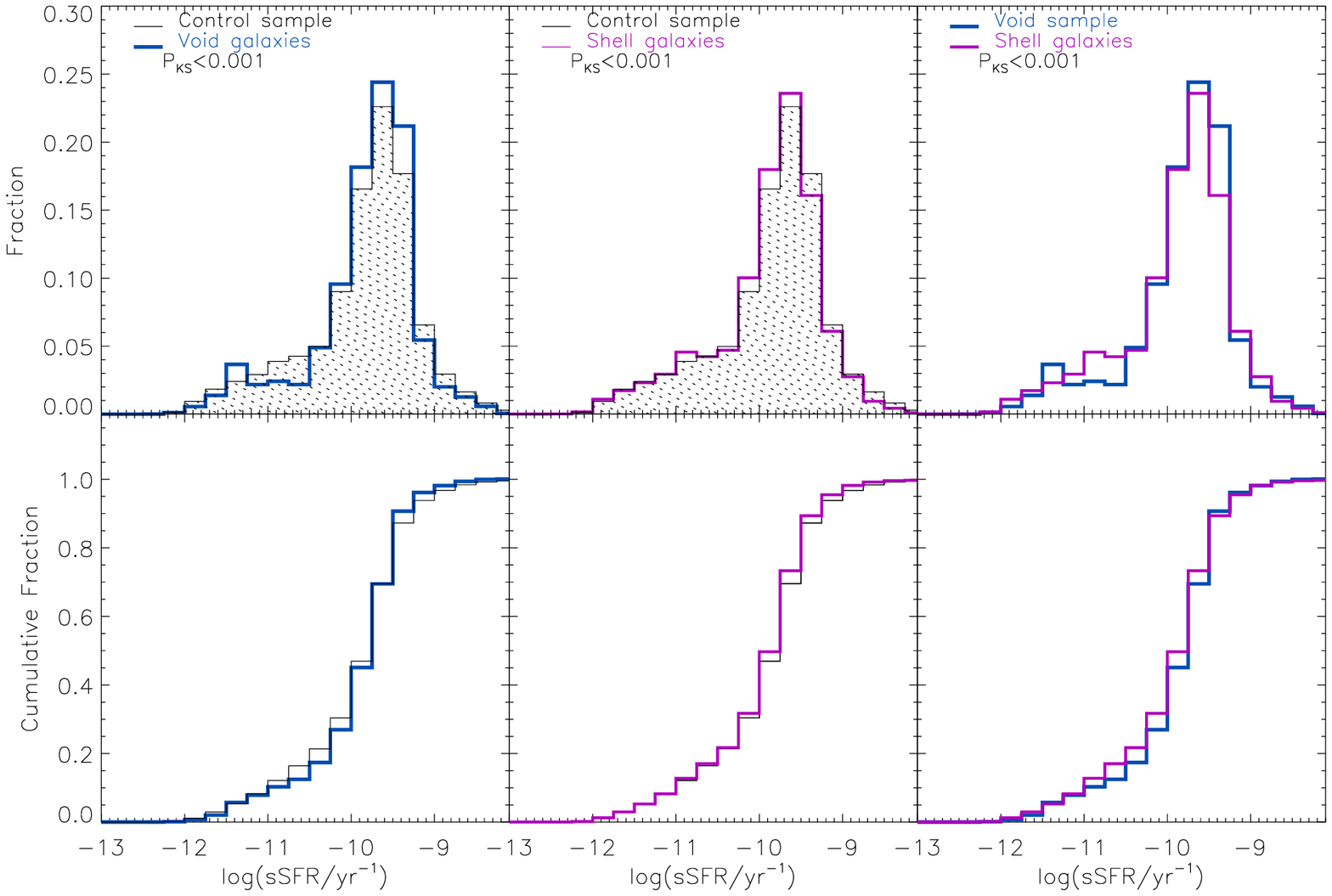}\\
\caption{Differential (upper panels) and cumulative (lower panels)
  sSFR distribution of the three samples,  using mass-matched samples
  for the void and shell galaxies. Colors are as in Fig. \ref{mass_hist}. }
\label{ssfr_hist_m} 
\end{figure*} 

One method to discriminate the environmental effect on galaxy
evolution is to compare the distribution of galaxy properties in the
different environments.
In Figs. \ref{mass_hist}-\ref{ssfr_hist} we show the distribution of
stellar mass, SFR and specific star formation rate, $sSFR=SFR/M_{*}$, 
of the three samples: voids, shells and control sample.  It is worth
to notice that the probability density functions (PDF) 
are calculated using the weights described in Sect. \ref{incompleteness}. Without such weighting scheme, the
contribution of low-mass objects would be extremely reduced. 
Here we focus the analysis of the shell sample only
on those galaxies lying at a void-centric distance $\leq
1.5 \cdot R_{void}$, in order to highlight any environmental effect with
respect to the control sample, that at larger distances is diluted.

\begin{figure*}
\includegraphics[width=2.2\columnwidth]{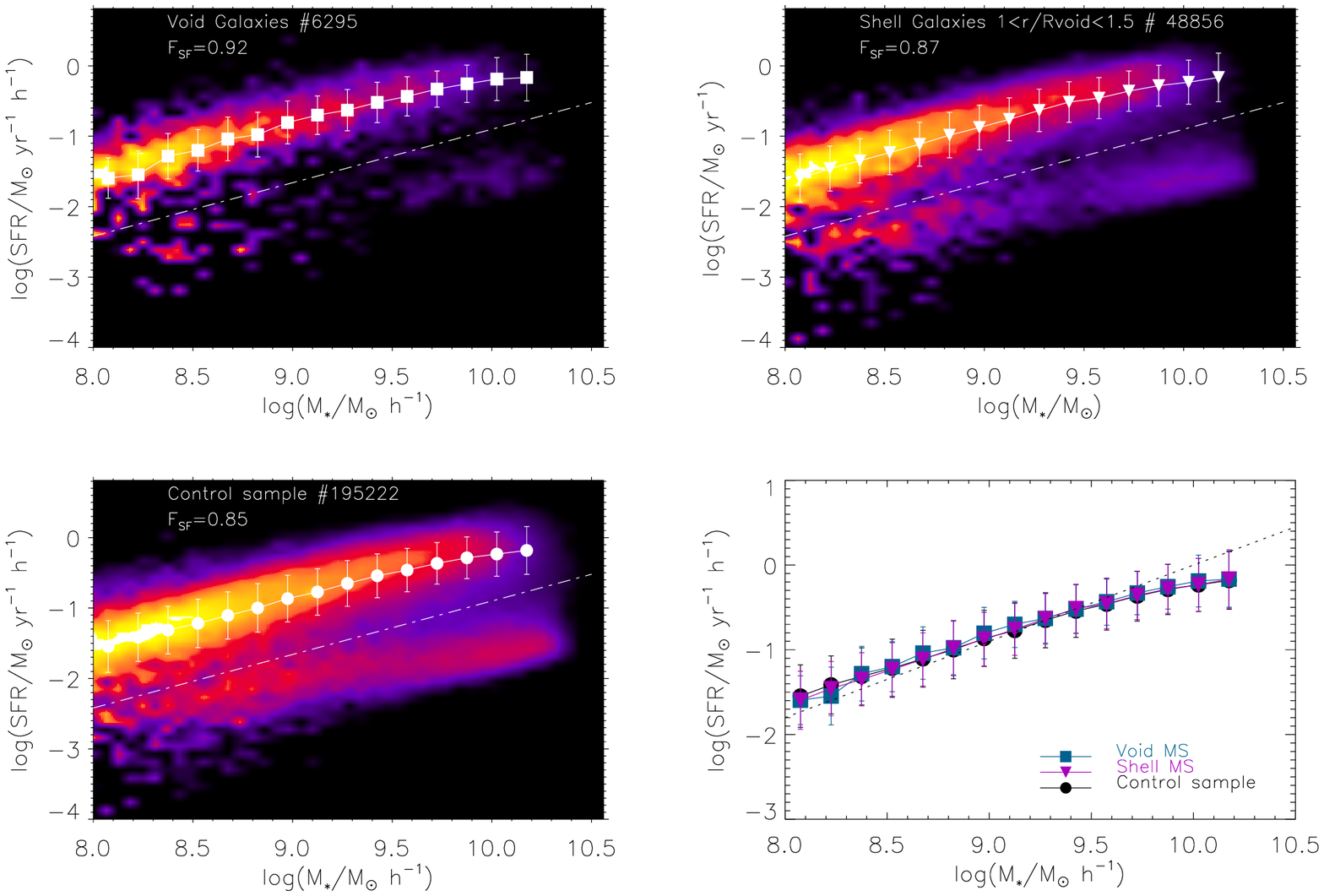}
\caption{SFR$-$stellar mass diagram for void galaxies (upper-left panel),
  shell galaxies with void-centric distance below $1.5\cdot R_{void}$
  (upper-right panel) and the control sample (lower-left
  panel).  Colors indicate the value of the two-dimensional weighted
  PDF, with yellow-white representing the highly populated regions and
  purple the poorly populated ones.  The dash-dotted line indicates
  the separation between the star forming and the
quiescent regions (Eq. \ref{sf_def}).  The mean SFR per mass bin for the star forming population
is shown by symbols with  error-bars, indicating the
standard deviation.
In the lower-right panel we report the MS of all three environments
for comparison. The dotted line is the MS
determination from \citet{Peng10}.}
\label{MS} 
\end{figure*} 

Voids have a mass distribution more abundant in low-mass galaxies then 
 shells and control sample (Fig. \ref{mass_hist}).  A  two-dimensional
 Kolmogorov-Smirnov (KS, \citealt{Peacock83}) test returns a low
 probability ($P<0.001$) that the two distributions are drawn from the same sample.
Control sample and shells display mass
 distribution apparently alike, although the KS test again rejects
 the null hypothesis that they are drawn from the same
 distribution. We notice that the same low probabilities are obtained
 also when using the Mann–Whitney U test \citep{MW74} in all the cases
 presented in this section.

Concerning the star formation properties, we see that low density
regions host galaxies with low star formation rates (Fig. \ref{sfr_hist}), a trend
that could be driven by the different mass distributions and the tight
correlation existing between SFR and stellar mass (see Sect. \ref{sect_MS}).
When focusing on the specific star formation rate
(Fig. \ref{ssfr_hist}), that gives a measure of the galaxy build-up time,
we see that void galaxies form stars
more efficiently, as their sSFRs
are significantly larger than those of shell and control galaxies.
The sSFRs of the shell galaxies are also
statistically larger than the control sample. In all the environments, we observe a strong peak at
sSFR$\sim 10^{-9.5} yr^{-1}$, representing the star forming
galaxies, and a tail at low sSFR representing the passive population,
that is not well represented in our sample.
Thus, the star forming and passive loci are found at similar sSFR values in all three environments, but the relative fraction of galaxies that populate the two loci varies between the environments.

Although  the star formation distributions in voids and shells appear
remarkably different than those in the control sample, we can not
exclude that this result is just a consequence of the mass bias. Indeed,
voids and shells host a larger number of low-mass
galaxies with respect to the control sample (see Fig.
\ref{mass_hist}), which may in principle bias the SFR and sSFR
distributions. 
To avoid the influence of the mass distribution, we have randomly extracted 10 subsamples from the void
and the shell samples, having the same mass distribution of the control sample.
In Figs. \ref{sfr_hist_m}-\ref{ssfr_hist_m} we show the SFR and sSFR
distributions for these mass-matched samples of voids and shells,
compared with the original control sample.
We do not show the mass distributions, as they are, by construction,
identical  to each other.  
The SFR distributions in this case appear remarkably similar, indicating that the low values of SFR in voids are mainly due to
the predominance of low mass galaxies.
However, the sSFR trend with environment is still
recovered, showing that the void galaxies have the passive population
strongly suppressed with respect to the control sample, whereas shell
and control galaxies have similar sSFR distributions.

\subsection{Main Sequence}\label{sect_MS}

A tight correlation between the star formation rate and the stellar
mass has been observed both in the local Universe \citep{Brinchmann04, Salim07, Peng10} and at high redshift 
\citep{Noeske07, Elbaz07, Rodighiero10, Rodighiero14}, widely known as
the Main Sequence (MS). 

In Fig. \ref{MS} we show how the galaxies of our samples are distributed
 in the SFR - stellar mass plane. The colors indicate the value of
the two-dimensional weighted probability distribution function (PDF). 
All the environments show two separate sequences: a well-defined main
sequence at high SFR and a cloud at low star formation formed by the quiescent
galaxies. These two sequences are analogous to the blue cloud and the red sequence 
observed in the colour-magnitude diagram \citep{Strateva01, Baldry04}. 
This bimodality motivates the division into star
forming and quiescent galaxies. 
In order to separate the two populations, we compute the SFR
distributions in small mass intervals. From the bimodal SFR distributions at
each mass bin, it is possible to define a minimum SFR, located in the valley between the
low- and high-SFR peaks. These minima give the
separating line between star forming and passive galaxies:
\begin{equation}\label{sf_def}
\log(SFR)=0.76 \log(M)-8.5 
\end{equation}
and it is shown by the dash-dotted line in Fig. \ref{MS}.  We have
applied this procedure to the control sample and used the same definition
of star forming galaxies for all the three samples. We notice
that, when translated into sSFR, our definition is in perfect agreement with
the one adopted by \citet{Omand14}.
All the galaxies lying above this line can be defined as star forming and used
to define the MS, given by the mean SFR in each stellar mass interval.
In the bottom-right panel of Fig. \ref{MS}  we show the comparison between the
MS in the three different environments, and we find that they are
completely consistent with each other and with the relation of Peng et
al. (2010), which was measured for the general environment with SDSS
data.
A similar scatter ($\sim 0.3
dex$) is also found for the three samples.

Although in rich clusters, the MS has been claimed to depend
on the local density \citep{Vulcani10}, in the field it has been
shown to be independent of the environment \citep{Peng10}.
Our results thus show that this 
is also valid down to the extreme low densities of cosmic voids (see
also \citealt{Kreckel12}).

\subsection{Galaxy properties vs void-centric distance}\label{dist}

\begin{figure}
\includegraphics[width=\columnwidth]{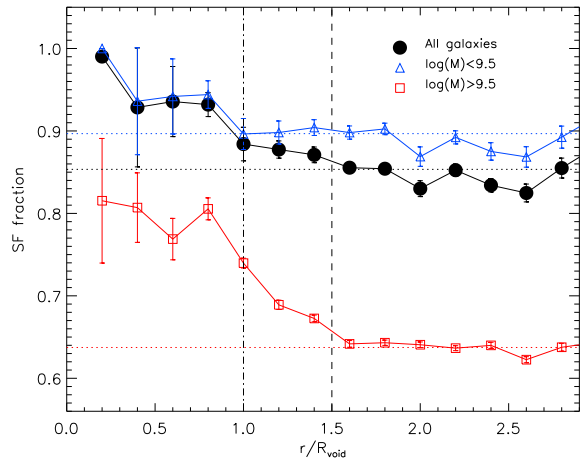}
\caption{Fraction of star forming galaxies as a function of
  void-centric distance for all (black circles), low-mass
  ($\log(M) \leq 9.5$, blue triangles) and high-mass galaxies
  ($\log(M>9.5$ red squares). Error-bars denote the 1$\sigma$ confidence
  intervals computed by means of 1000 bootstrap resamplings.
  The horizontal lines indicate the star
  forming fraction in the general sample. The vertical lines indicate
  the position of the void edge (dot-dashed line) and the value:
  $r/R_{void}=1.5$ where the shell properties converge to those of the
  control sample (dashed line).}
\label{FRC_dist} 
\end{figure}

In this section we investigate how galaxy properties depend on the galaxy
position within the void. 
In Fig. \ref{FRC_dist} we show how the fraction of star forming
galaxies depends on the void-centric distance, normalized to the
void size. Star forming galaxies are defined as those galaxies lying
above the line defined in Eq. \ref{sf_def}.  
The confidence interval has been estimated by means of 1000 bootstrap resamplings.
We find that the fraction of star forming galaxies in the innermost part of the
void is close to 1 and then it steadily decreases as
moving to large distances from the void center. Interestingly, the
star forming fraction converges to the mean value measured for the control sample for
distance $r\geq 1.5 \cdot R_{void}$. 
When splitting the sample in low- and
high-mass galaxies, we still observe a significant decrease in the
star forming fraction in the two sub-samples, indicating that the
trend is not simply driven by an increase of the mean galaxy mass towards the void
edge. In the low-mass sample, the decaying appears milder, as the vast
majority of low-mass galaxies are star forming  in all the
environments. The convergence  to the star forming fraction of the
control value for $r\geq 1.5 \cdot R_{void}$  holds also for the two sub-samples.

\begin{figure}
\includegraphics[width=\columnwidth]{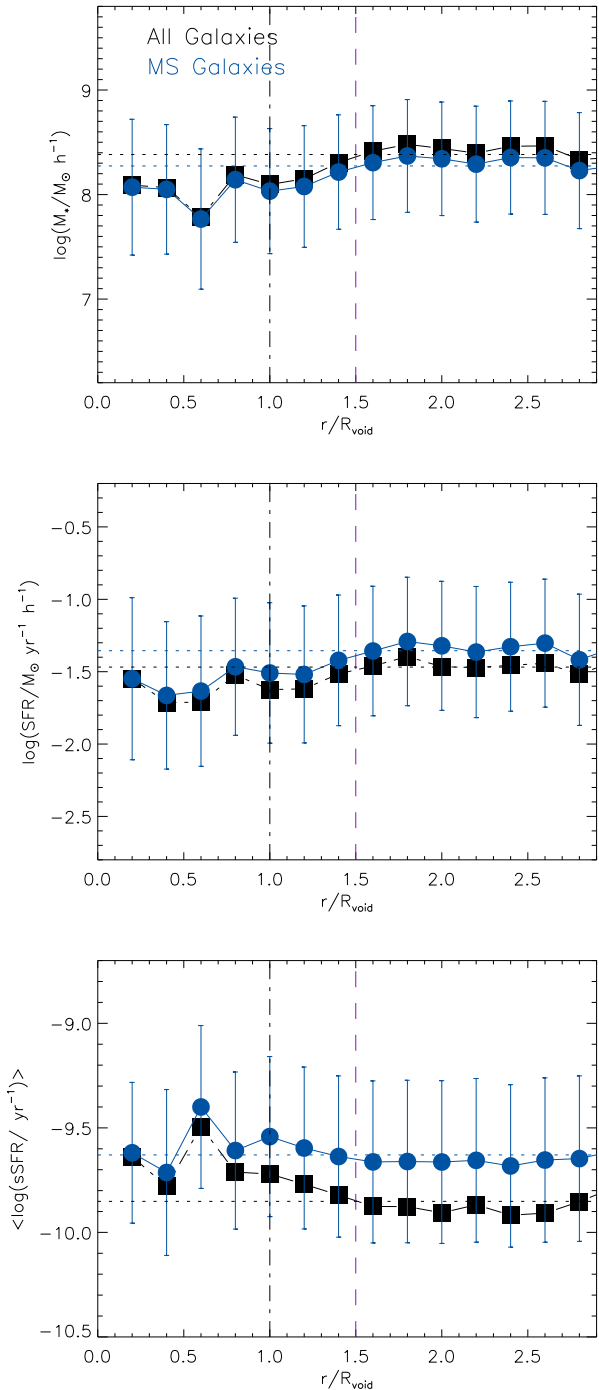}
\caption{Galaxy properties as a function of void-centric
  distance. From top to bottom we show: stellar mass, SFR and
  sSFR. Black squares indicate the mean values for the whole sample,
  whereas blue circles stay for the main sequence
  galaxies. For clearness we only show the standard deviations for the
  MS samples. Horizontal dotted lines denote the mean quantities in the
control sample for all the galaxies (black) and the MS galaxies
(blue). Vertical lines indicate the positions of the void edge (black dot-dashed)
and the shell limit (purple dashed).}
\label{SFR_dist} 
\end{figure}

In Fig. \ref{SFR_dist} we show  how galaxy properties, namely
$M_{*}$, SFR and sSFR, change with the  void-centric distance.
We find lower stellar masses and lower SFRs in
  the central region of the void, although the errorbars are too large
  to draw robust conclusion. On the other hand,  the sSFR of the main
  sample appears to decrease at large distances,
pointing towards a progressive reduction of the star formation activity of
the global sample.  Although the trend of the sSFR with respect to the
void-centric distance appears in contradiction with the behavior of
the SFR that increases at large distances, the latter quantity is well
correlated with the stellar mass (see Sect. 3.2) which is driving the
SFR trend with void-centric distance.

The progressive reduction in the star formation activity with the
void-centric distance can be interpreted as a manifestation of the
SFR-density relation within void regions, ought to an increasing local
density approaching the void edge, as shown by the void density profiles
\citep{Colberg05, Ricciardelli13, Ricciardelli14}. The convergence of
galaxy properties to the control sample value allows us to define the region of
influence of voids as  extending up to $1.5 \cdot R_{voids}$.

\subsection{Galaxy properties vs void size}\label{dist_rvoid}

\begin{figure}
\includegraphics[width=\columnwidth]{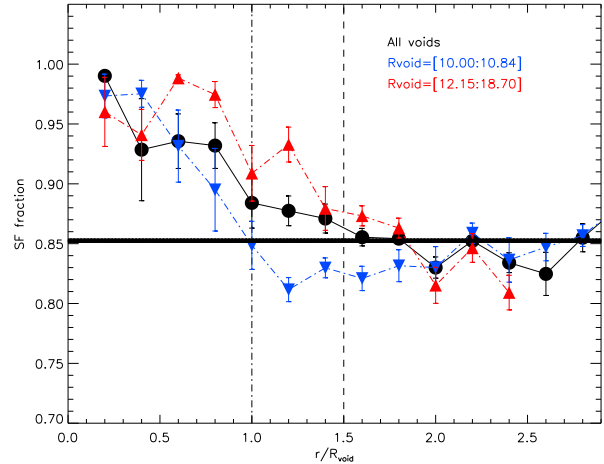}
\caption{Fraction of star forming galaxies as a function of
  void-centric distance for voids of different size. The black symbols
indicate the whole sample of voids, the blue ones refer to voids in
the lowest quartile of the size distribution and the red ones to those in the upper
quartile. }
\label{FRC_rvoid} 
\end{figure} 

In this section, we explore whether the star formation activity in
voids and void shells depends on the size of the voids.
The void sample has been divided  into four equally-populated subsamples at
different radii, containing $\sim175$ voids each. 
The star formation fraction as a function of the
void-centric distance for the two extreme quartiles at small and large
void radii is shown in Fig. \ref{FRC_rvoid} and compared with the
curve of the whole void sample. 
In the inner part of the voids, the small and large voids do not show
significant differences. However, when approaching the void edge  ($0.8 \lesssim r/R_{void} \lesssim
1.5$), the two curves start to deviate, with the
large voids showing an enhanced star formation activity in their
shells with respect to their smaller counterparts. 
Although the low number statistics prevent us to draw robust
conclusions, this result would support the theoretical expectations
that small and large voids have different dynamical evolution, that is
determined by the large scale region surrounding them
\citep{Sheth04}.  In support of this, there are evidences that small
voids are surrounded by an overdense shell, that in the larger voids
is not observed (\citealt{Ceccarelli13}, \citealt{Hamaus14}, but see
also \citealt{Nadathur14}).

\section{Conclusions}\label{conclusions}

In this work we have performed a statistical study of the star
formation activity in voids and void shells and compared it with a
control galaxy sample. 
As previously observed, we find that void galaxies are characterized by lower stellar masses and 
 enhanced star formation activity with respect to galaxies in the
 general environment. Shells appear as a transition region, having
 galaxies with intermediate properties between void galaxies and the general galaxy
 population. We can exclude that the
difference in the sSFR of the three environments under consideration
is due to the different mass distributions, as a similar result on the
sSFR distribution is obtained when comparing mass-matched sub-samples.

The effect of void environment manifests itself in the different
proportion of star forming and passive galaxies, but the average sSFR 
of the two populations taken separately does not depend on the large-scale
environment.  This result is more clearly observed in the star forming
main sequence, that is remarkably constant over the three
environments.  That is to say, when the stellar mass and the
population type (star forming or passive) are fixed, the star formation activity does not
depend on the environment. 
This result is consistent with the findings in the general field
indicating that there is no strong effect due to local density on the
star forming MS \citep{Peng10, Wijesinghe12}. 
This uniformity  of the MS over the large-scale environment 
 would indicate that environmental effects are fast-acting, thus
manifesting themselves as quenching mechanisms, that
rapidly move the galaxy from the MS to the passive region.
Conversely,  the regulation of the supply of gas available for star
formation in those galaxies that are not quenched, depends largely  on the
mass of the host halo and has no further relation with environment. 

We also find a significant radial variation of the star forming
fraction, with the innermost part of voids having the passive
population strongly suppressed.   Interestingly, the properties related
to the void-centric distance converge to the general field values for
a fixed radius of $\sim1.5r/R_{void}$, that can serve as suitable
definition of void shell.
Furthermore, we observe a different
star formation activity in the shells of voids of different sizes, with
the small voids having a more abundant passive population with respect
to the larger voids. This  could be related to the different dynamical
evolution experienced by voids of different sizes \citep{Sheth04, Ceccarelli13}.

One aspect that still needs to be understood is whether the
enhancement in the star formation activity within voids is just a
consequence of the SFR-density relation, being the inner regions of
voids extremely rarefied (see Ricciardelli et al. 2014), or there is an
additional influence of voids beside that of the local density (see
Ceccarelli et al. 2008). We plan to test this issue in a future work.

\section*{Acknowledgements} 

This work was  supported by the Spanish Ministerio de Econom\'{\i}a y Competitividad 
(MINECO, grants   AYA2010-21322-C03-01) and    the   Generalitat   Valenciana   (grant
PROMETEO-2009-103).
The work of AC is supported by the STARFORM Sinergia Project funded by
the Swiss National Science Foundation. 
 J.V. did part of the work thanks to a post-doc
fellowship from the former Spanish Ministry of Science and Innovation
under programs 3I2005 and 3I2406. J.V. also acknowledges the financial
support from the FITE (Fondos de In
versi\'{o}n de Teruel) and the Spanish grant AYA2012-30789.
 ER acknowledges the kind
hospitality of the Observatoire de Gen{\`e}ve (Universit{\'e} de Gen{\`e}ve)
during the preparation of part of this work.

This work has used data from SDSS Data Release 7. Funding for the SDSS and SDSS-II has been provided by the Alfred P. Sloan Foundation, the Participating Institutions, the National Science Foundation, the U.S. Department of Energy, the National Aeronautics and Space Administration, the Japanese Monbukagakusho, the Max Planck Society, and the Higher Education Funding Council for England. The SDSS Web Site is http://www.sdss.org/.


\newpage

\begin{thebibliography}{}



\bibitem[\protect\citeauthoryear{Alpaslan et 
al.}{2014}]{Alpaslan14} Alpaslan M., et al., 2014, MNRAS, 440, 
L106 

\bibitem[\protect\citeauthoryear{Arag{\'o}n-Calvo et 
al.}{2010}]{AragonCalvo10} Arag{\'o}n-Calvo M.~A., Platen E., van de 
Weygaert R., Szalay A.~S., 2010, ApJ, 723, 364 

\bibitem[\protect\citeauthoryear{Aragon-Calvo 
\& Szalay}{2013}]{AragonCalvo13} Aragon-Calvo M.~A., Szalay A.~S., 2013, MNRAS, 428, 3409 

\bibitem[\protect\citeauthoryear{Baldry et al.}{2004}]{Baldry04} 
Baldry I.~K., Glazebrook K., Brinkmann J., Ivezi{\'c} {\v Z}., Lupton 
R.~H., Nichol R.~C., Szalay A.~S., 2004, ApJ, 600, 681 


\bibitem[\protect\citeauthoryear{Baldwin, Phillips, 
\& Terlevich}{1981}]{Baldwin81} Baldwin J.~A., Phillips M.~M., Terlevich R., 1981, PASP, 93, 5 

\bibitem[\protect\citeauthoryear{Balogh et al.}{1997}]{Balogh97} 
Balogh M.~L., Morris S.~L., Yee H.~K.~C., Carlberg R.~G., Ellingson E., 
1997, ApJ, 488, L75 

\bibitem[\protect\citeauthoryear{Beygu et al.}{2013}]{Beygu13} 
Beygu B., Kreckel K., van de Weygaert R., van der Hulst J.~M., van Gorkom 
J.~H., 2013, AJ, 145, 120 

\bibitem[\protect\citeauthoryear{Blanton et 
al.}{2005}]{Blanton05} Blanton M.~R., et al., 2005, AJ, 129, 2562 

\bibitem[\protect\citeauthoryear{Blanton 
\& Roweis}{2007}]{Blanton07} Blanton M.~R., Roweis S., 2007, AJ, 133, 734 

\bibitem[\protect\citeauthoryear{Bond, Kofman, 
\& Pogosyan}{1996}]{Bond96} Bond J.~R., Kofman L., Pogosyan D., 1996,
Nature, 380, 603


\bibitem[\protect\citeauthoryear{Brinchmann et 
al.}{2004}]{Brinchmann04} Brinchmann J., Charlot S., White S.~D.~M., 
Tremonti C., Kauffmann G., Heckman T., Brinkmann J., 2004, MNRAS, 351,
1151

\bibitem[\protect\citeauthoryear{Bruzual 
\& Charlot}{2003}]{BC03} Bruzual G., Charlot S., 2003, MNRAS, 344, 1000 


\bibitem[\protect\citeauthoryear{Charlot 
\& Fall}{2000}]{2000ApJ...539..718C} Charlot S., Fall S.~M., 2000, ApJ, 539, 718 


\bibitem[\protect\citeauthoryear{Cautun et al.}{2014}]{Cautun14} 
Cautun M., van de Weygaert R., Jones B.~J.~T., Frenk C.~S., 2014, MNRAS, 
441, 2923 

\bibitem[\protect\citeauthoryear{Ceccarelli et 
al.}{2013}]{Ceccarelli13} Ceccarelli L., Paz D., Lares M., Padilla 
N., Lambas D.~G., 2013, MNRAS, 434, 1435 


\bibitem[\protect\citeauthoryear{Ceccarelli, Padilla, 
\& Lambas}{2008}]{Ceccarelli08} Ceccarelli L., Padilla N., Lambas D.~G., 2008, MNRAS, 390, L9 


\bibitem[\protect\citeauthoryear{Charlot 
\& Fall}{2000}]{CF00} Charlot S., Fall S.~M., 2000,
ApJ, 539, 718 

\bibitem[\protect\citeauthoryear{Charlot 
\& Longhetti}{2001}]{CL01} Charlot S., Longhetti M., 2001, MNRAS, 323, 887 



\bibitem[\protect\citeauthoryear{Colberg et 
al.}{2005}]{Colberg05} Colberg J.~M., Sheth R.~K., Diaferio A., 
Gao L., Yoshida N., 2005, MNRAS, 360, 216 

\bibitem[\protect\citeauthoryear{Colless et 
al.}{2001}]{Colless01} Colless M., et al., 2001, MNRAS, 328, 1039 

\bibitem[\protect\citeauthoryear{Cybulski et 
al.}{2014}]{Cybulski14} Cybulski R., Yun M.~S., Fazio G.~G., 
Gutermuth R.~A., 2014, MNRAS, 439, 3564 


\bibitem[\protect\citeauthoryear{Dressler}{1980}]{Dressler80} 
Dressler A., 1980, ApJ, 236, 351 

\bibitem[\protect\citeauthoryear{Elbaz et 
al.}{2007}]{Elbaz07} Elbaz D., et al., 2007, A\&A, 468, 33 

\bibitem[\protect\citeauthoryear{Gregory 
\& Thompson}{1978}]{Gregory78} Gregory S.~A., Thompson L.~A., 1978, ApJ, 222, 784 



\bibitem[\protect\citeauthoryear{Hamaus, Sutter, 
\& Wandelt}{2014}]{Hamaus14} Hamaus N., Sutter P.~M., Wandelt B.~D., 2014, PhRvL, 112, 251302 


\bibitem[\protect\citeauthoryear{Hoyle, Vogeley, 
\& Pan}{2012}]{Hoyle12} Hoyle F., Vogeley M.~S., Pan D., 2012, MNRAS, 426, 3041 

\bibitem[\protect\citeauthoryear{Kauffmann et 
al.}{2003}]{Kauff03} Kauffmann G., et al., 2003, MNRAS, 341, 33 

\bibitem[\protect\citeauthoryear{Kirshner et 
al.}{1981}]{Kirshner81} Kirshner R.~P., Oemler A., Jr., Schechter 
P.~L., Shectman S.~A., 1981, ApJ, 248, L57 

\bibitem[\protect\citeauthoryear{Kroupa}{2001}]{Kroupa01} Kroupa 
P., 2001, MNRAS, 322, 231 


\bibitem[\protect\citeauthoryear{Kreckel et 
al.}{2012}]{Kreckel12} Kreckel K., Platen E., Arag{\'o}n-Calvo 
M.~A., van Gorkom J.~H., van de Weygaert R., van der Hulst J.~M., Beygu B., 
2012, AJ, 144, 16 





\bibitem[\protect\citeauthoryear{Mann \& Whitney}{1974}]{MW74} Mann H. B., Whitney D. R., 1947, Ann. Math. Statistics, 18, 50

\bibitem[\protect\citeauthoryear{Nadathur 
\& Hotchkiss}{2013}]{Nadathur13} Nadathur S., Hotchkiss S., 2013, arXiv, arXiv:1310.2791 

\bibitem[\protect\citeauthoryear{Nadathur et 
al.}{2014}]{Nadathur14} Nadathur S., Hotchkiss S., Diego J.~M., 
Iliev I.~T., Gottl{\"o}ber S., Watson W.~A., Yepes G., 2014, arXiv, 
arXiv:1407.1295 


\bibitem[\protect\citeauthoryear{Noeske et al.}{2007}]{Noeske07} 
Noeske K.~G., et al., 2007, ApJ, 660, L47 

\bibitem[\protect\citeauthoryear{Omand, Balogh, 
\& Poggianti}{2014}]{Omand14} Omand C.~M.~B., Balogh M.~L., Poggianti B.~M., 2014, MNRAS, 440, 843 

\bibitem[\protect\citeauthoryear{Pan et al.}{2012}]{Pan12} 
Pan D.~C., Vogeley M.~S., Hoyle F., Choi Y.-Y., Park C., 2012, MNRAS, 421, 
926 

\bibitem[\protect\citeauthoryear{Peng et al.}{2010}]{Peng10} 
Peng Y.-j., et al., 2010, ApJ, 721, 193 


\bibitem[\protect\citeauthoryear{Patiri et al.}{2006}]{Patiri06} 
Patiri S.~G., Prada F., Holtzman J., Klypin A., Betancort-Rijo J., 2006, 
MNRAS, 372, 1710

\bibitem[\protect\citeauthoryear{Peacock}{1983}]{Peacock83} 
Peacock J.~A., 1983, MNRAS, 202, 615 

\bibitem[\protect\citeauthoryear{Poggianti et 
al.}{1999}]{Poggianti99} Poggianti B.~M., Smail I., Dressler A., 
Couch W.~J., Barger A.~J., Butcher H., Ellis R.~S., Oemler A., Jr., 1999, 
ApJ, 518, 576 

\bibitem[\protect\citeauthoryear{Pustilnik et 
al.}{2013}]{Pustilnik13} Pustilnik S.~A., Martin J.-M., Lyamina 
Y.~A., Kniazev A.~Y., 2013, MNRAS, 432, 2224 


\bibitem[\protect\citeauthoryear{Pustilnik et 
al.}{2011}]{Pustilnik11} Pustilnik S.~A., Martin J.-M., Tepliakova 
A.~L., Kniazev A.~Y., 2011, MNRAS, 417, 1335 





\bibitem[\protect\citeauthoryear{Ricciardelli, Quilis, 
\& Planelles}{2013}]{Ricciardelli13} Ricciardelli E., Quilis V.,
Planelles S., 2013, MNRAS, 434, 1192  

\bibitem[\protect\citeauthoryear{Ricciardelli, Quilis, 
\& Varela}{2014}]{Ricciardelli14} Ricciardelli E., Quilis V., Varela J., 2014, MNRAS, 440, 601 

\bibitem[\protect\citeauthoryear{Rieder et al.}{2013}]{Rieder13} 
Rieder S., van de Weygaert R., Cautun M., Beygu B., Portegies Zwart S., 
2013, MNRAS, 435, 222 

\bibitem[\protect\citeauthoryear{Rodighiero et 
al.}{2014}]{Rodighiero14} Rodighiero G., et al., 2014, arXiv, 
arXiv:1406.1189 

\bibitem[\protect\citeauthoryear{Rodighiero et 
al.}{2010}]{Rodighiero10} Rodighiero G., et al., 2010, A\&A, 518, L25 

\bibitem[\protect\citeauthoryear{Rojas et al.}{2005}]{Rojas05} 
Rojas R.~R., Vogeley M.~S., Hoyle F., Brinkmann J., 2005, ApJ, 624, 571 

\bibitem[\protect\citeauthoryear{Rojas et al.}{2004}]{Rojas04} 
Rojas R.~R., Vogeley M.~S., Hoyle F., Brinkmann J., 2004, ApJ, 617, 50 

\bibitem[\protect\citeauthoryear{Salim et al.}{2007}]{Salim07} 
Salim S., et al., 2007, ApJS, 173, 267 

\bibitem[\protect\citeauthoryear{Sheth 
\& van de Weygaert}{2004}]{Sheth04} Sheth R.~K., van de Weygaert R., 2004, MNRAS, 350, 517 

\bibitem[\protect\citeauthoryear{Strateva et 
al.}{2001}]{Strateva01} Strateva I., et al., 2001, AJ, 122,
1861 






\bibitem[\protect\citeauthoryear{York et al.}{2000}]{York00} 
York D.~G., et al., 2000, AJ, 120, 1579 

\bibitem[\protect\citeauthoryear{van den Bosch et 
al.}{2008}]{vdb08} van den Bosch F.~C., Aquino D., Yang X., 
Mo H.~J., Pasquali A., McIntosh D.~H., Weinmann S.~M., Kang X., 2008, 
MNRAS, 387, 79 


\bibitem[\protect\citeauthoryear{van de Weygaert 
\& Platen}{2011}]{Vandew11} van de Weygaert R., Platen E., 2011, IJMPS, 1, 41 

\bibitem[\protect\citeauthoryear{van de Weygaert 
\& van Kampen}{1993}]{Vandew93} van de Weygaert R., van Kampen E., 1993, MNRAS, 263, 481 

\bibitem[\protect\citeauthoryear{Varela et al.}{2012}]{Varela12} 
Varela J., Betancort-Rijo J., Trujillo I., Ricciardelli E., 2012, ApJ, 744, 
82

\bibitem[\protect\citeauthoryear{von Benda-Beckmann \& M{\"u}ller}{2008}]{vonbenda08} von Benda-Beckmann A.~M., M{\"u}ller V., 2008, MNRAS, 384, 1189 

\bibitem[\protect\citeauthoryear{Vulcani et 
al.}{2010}]{Vulcani10} Vulcani B., Poggianti B.~M., Finn R.~A., 
Rudnick G., Desai V., Bamford S., 2010, ApJ, 710, L1 

\bibitem[\protect\citeauthoryear{Wijesinghe et 
al.}{2012}]{Wijesinghe12} Wijesinghe D.~B., et al., 2012, MNRAS, 
423, 3679 

\end{thebibliography}
\end{document}